\documentclass[prd,twocolumn,superscriptaddress,nofootinbib]{revtex4-2}
\usepackage{url}

\usepackage{float}
\usepackage{epsfig}
\usepackage{amsfonts}
\usepackage{graphicx}
\usepackage{amsmath}
\usepackage{amssymb}
\usepackage{color}
\usepackage{bbm}
\usepackage{dcolumn}
\usepackage{bm}
\usepackage{ulem}
\usepackage{mathrsfs}
\usepackage{bbold}
\usepackage{datetime}
\usepackage{nameref}

\usepackage{rotating}
\usepackage[usenames,dvipsnames]{xcolor}
\usepackage[colorlinks=true,citecolor=blue,linkcolor=blue,urlcolor=Green]{hyperref}
\usepackage{lipsum} 
\usepackage{tikz,tikz-3dplot}
\usepackage{etoolbox} 
\usepackage[capitalize]{cleveref}
\usepackage{extarrows}

\def\bc{\begin{center}}
\def\nno{\nonumber}
\def\ec{\end{center}}
\def\be{\begin{eqnarray}}
\def\ee{\end{eqnarray}}
\definecolor{dyellow}{rgb}{1.,0.8,.0}
\definecolor{myblue}{rgb}{.1,.1,.7}
\definecolor{dcyan}{rgb}{.0,.6,.6}
\definecolor{dmagenta}{rgb}{0.6,0.0,0.6}
\definecolor{brown}{rgb}{0.6,0.2,0.}
\definecolor{darkblue}{rgb}{.0,.0,0.5}
\definecolor{darkred}{rgb}{0.75,0.0,0.0}
\definecolor{orange}{rgb}{1.,.6,.0}
\definecolor{dorange}{rgb}{0.8,.4,.0}
\definecolor{darkgreen}{rgb}{0.0,0.6,0.0}
\definecolor{purple}{rgb}{.4,.0,.4}
\definecolor{lightgrey}{rgb}{0.7, 0.7, 0.7}
\definecolor{grey}{rgb}{0.4, 0.4, 0.4}

\usepackage{geometry}
\usepackage[font=small,labelfont=bf,justification=raggedright]{caption}

\newcommand{\xdownarrow}[1]{%
  {\left\downarrow\vbox to #1{}\right.\kern-\nulldelimiterspace}
}
\newcommand{\xuparrow}[1]{%
  {\left\uparrow\vbox to #1{}\right.\kern-\nulldelimiterspace}
}

\definecolor{myred}{RGB}{189, 38, 49}

\begin{document}
\title{A universal approach to Renyi entropy of multiple disjoint intervals}
\author{Han-Qing Shi} \email{by2030104@buaa.edu.cn}
\affiliation{Center for Gravitational Physics, Department of Space Science, Beihang University, Beijing 100191, China}
\author{Hai-Qing Zhang} \email{hqzhang@buaa.edu.cn}
\affiliation{Center for Gravitational Physics, Department of Space Science, Beihang University, Beijing 100191, China}
\affiliation{Peng Huanwu Collaborative Center for Research and Education, Beihang University, Beijing 100191, China}

\begin{abstract}
We develop a general theory for computing the Renyi entropy with general multiple disjoint intervals from the swapping operations. Our theory is proposed based on the fact that we have observed the resemblance between the replica trick in quantum field theory and the swapping operation. Consequently, the Renyi entropy can be obtained by evaluating the expectation values of the swapping operator. As an application, we study the Renyi entropy of a one-dimensional transverse-field Ising model for two, three and four disjoint intervals. As the system is at the critical point, our computations of the Renyi entropy are consistent with the analytical results from the conformal field theory. Moreover, our methods can go beyond the critical regime of the Ising model.   
\end{abstract}

\maketitle


\section{Introduction}


Entanglement is one of the most mysterious phenomena in quantum physics \cite{horodecki2009quantum}. The entanglement entropy (von Neumann entropy) can be used to quantify how the systems are entangled \cite{nielsen2010quantum}. Recently, Renyi entropy $S_m$, a generalization of the von Neumann entropy, has attracted much attention \cite{renyi1961measures}. In the limit of $m\to1$, the Renyi entropy $S_m$ returns to the von Neumann entropy. Moreover, Renyi entropy has wider applications in quantum physics since it encodes much more information than the von Neumann entropy, for instance in studying the entanglement spectrum of the density matrix \cite{PhysRevLett.101.010504}. In a realistic system, analytically computing the von Neumann entropy or the Renyi entropy is a formidable task due to the large Hilbert space \cite{muller2013quantum}. Only in some special cases, such as in conformal field theory, the entanglement entropies can be studied analytically \cite{calabrese2009entanglement}. Renyi entropy or entanglement entropy for multiparty subsystems has been studied extensively in recent years, ranging from the condensed matter physics \cite{alba2010entanglement,fagotti2010entanglement,calabrese2009entanglement2,calabrese2011entanglement} to AdS/CFT correspondences \cite{headrick2010entanglement,faulkner2013entanglement}. Experiments towards this direction were also performed in \cite{xiong2022scaling}. 

However, these literatures were mainly dealing with two disjoint subsystems, for general multiple disjoint subsystems the computation will be more involved. In conformal field theory, although Renyi entropy or entanglement entropy can still be calculated for subsystems with two disjoint intervals, it requires four-point correlation functions and a model dependent universal function \cite{alba2010entanglement,furukawa102mutual}. For more disjoint intervals, higher-order correlation functions are needed which makes the computation more involved. Away from the conformal field theory, there are still no closed formula to study the entropy for multiple disjoint systems. Therefore, it is necessary to develop a new method for calculating the Renyi entropy or entanglement entropy of arbitrary disjoint subsystems.

In this paper, we give a systematic approach to analytically compute the Renyi entropy $S_m$ for general multiple disjoint intervals. In particular, we adopt the `swapping operations' to compute the Renyi entropy. The details of the swapping operation can be found in the main text. This swapping operation was previously used to compute the second Renyi entropy $S_2$ for one disjoint subsystem \cite{hastings2010measuring}. In quantum field theory, people usually adopt the `replica trick' in the path integral to compute the Renyi entropy \cite{nishioka2009holographic}. By observing the resemblance between the swapping operation and the replica trick in quantum field theory, we propose and generalize the swapping operation to compute the general Renyi entropy $S_m$ for general multiple disjoint intervals. As a result, the Renyi entropy is equivalent to a logarithmic function of the expectation values of the swapping operator. As an application of our proposal, we study the Renyi entropy $S_m$ for multiple disjoint intervals in the ground state of the transverse-field quantum Ising model (TFQIM). In particular, we study the second Renyi entropy $S_2$ with two, three and four disjoint intervals for various transverse magnetic fields. At the critical point of phase transition from paramagnetic to ferromagnetic phase, we find that our results for two disjoint intervals are consistent with those from the conformal field theory \cite{alba2010entanglement,furukawa102mutual}. Moreover, our methods can be applied in non-critical regimes which cannot be described by conformal field theory or by analytic closed formula. Therefore, in principle, our methods for the generalized swapping operation can be used to study any order of the Renyi entropy $S_m$ for any disjoint intervals.   


\section{Theory}

\subsection{Second-order Renyi entropy $S_2$ for two disjoint intervals}
The $m$-th order Renyi entropy is defined as \cite{renyi1961measures},
\begin{equation}
	S_m\equiv\frac{1}{1-m}\ln {\rm Tr}(\rho_A^m)
\end{equation}	
where $\rho_A$ is the reduced density matrix of a subsystem $A$, obtained by tracing out the whole density matrix over the complement of $A$. 

\begin{figure}[t]
	\centering
	\includegraphics[trim=4.5cm 3.5cm 5.cm 2.5cm, clip=true, scale=0.8]{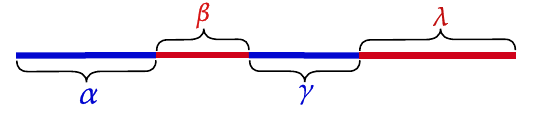}
	\caption{The whole interval is divided into four parts, $\alpha$, $\beta$, $\gamma$ and $\lambda$. The subsystem $A$ consists of two disjoint parts, i.e., $A=\alpha\cup\gamma$, while the complement is $B=\beta\cup\lambda$. }
	\label{fig1}
\end{figure}

For a better understanding, let's first consider the second order Renyi entropy $S_2$ with two disjoint intervals $S_2=-\ln {\rm Tr}(\rho_A^2)$. In this case we need to divide the whole system into four parts, see Fig.\ref{fig1} for illustration. The subsystem $A$ (blue parts) consists of two disjoint intervals, i.e., $A=\alpha\cup\gamma$, while the complement part is $B=\bar A=\beta\cup\lambda$. Assume that the state of the whole system is $|\Psi\rangle$, which can be decomposed as the superpositions of each subsystem, 
\begin{equation}
	|\Psi\rangle=\sum_{\alpha\beta\gamma\lambda}C_{\alpha\beta\gamma\lambda}|\alpha\rangle|\beta\rangle|\gamma\rangle|\lambda\rangle,
\end{equation}
where $C_{\alpha\beta\gamma\lambda}$ is the coefficient while $|\alpha\rangle, |\beta\rangle, |\gamma\rangle$ and $|\lambda\rangle$ represent the basis of each subsystem, respectively. Then, we can build another auxiliary state $|\Psi\rangle$ which is exactly the copy of the existing $|\Psi\rangle$. The two can form a composite state $|\Psi\rangle\otimes|\Psi\rangle$. Therefore, we can define a unitary swapping operator $S^{(2)}_{\rm wap}$ which only acts on the subsystem $A$ by exchanging the states $|\alpha_1\rangle\leftrightarrow|\alpha_2\rangle$ and $|\gamma_1\rangle\leftrightarrow|\gamma_2\rangle$ coming from the two states $|\Psi\rangle$. The superscript index $(2)$ over $S^{(2)}_{\rm wap}$ indicates that it will exchange the states in the two $|\Psi\rangle$'s.  The computational rule is as follows, \footnote{The swapping operator needs to act on two subsystem states, one comes from the original system and the other one comes from the copy of the system. The detailed computations can be found in the \hyperref[appA]{Appendix A}. } 
\begin{widetext}
\begin{eqnarray}\label{swap1}
	S^{(2)}_{\rm wap}|\Psi\rangle\otimes|\Psi\rangle&=&S^{(2)}_{\rm wap}\left(\sum_1 C_{\alpha_1\beta_1\gamma_1\lambda_1}|\alpha_1\rangle|\beta_1\rangle|\gamma_1\rangle|\lambda_1\rangle\right)\otimes\left(\sum_2 C_{\alpha_2\beta_2\gamma_2\lambda_2}|\alpha_2\rangle|\beta_2\rangle|\gamma_2\rangle|\lambda_2\rangle\right)\nonumber\\
	&=&\sum_{12}C_{\alpha_1\beta_1\gamma_1\lambda_1}C_{\alpha_2\beta_2\gamma_2\lambda_2}|\alpha_2\rangle|\beta_1\rangle|\gamma_2\rangle|\lambda_1\rangle\otimes|\alpha_1\rangle|\beta_2\rangle|\gamma_1\rangle|\lambda_2\rangle,
\end{eqnarray}
\end{widetext}
in which $\sum_i\equiv\sum_{\alpha_i\beta_i\gamma_i\lambda_i}$ with $i=1, 2$. Therefore, the expectation value of the swapping operator can be evaluated as, 
\begin{eqnarray}\label{s2}
	&&\langle S^{(2)}_{\rm wap}\rangle\equiv\langle\Psi\otimes\Psi|S^{(2)}_{\rm wap}|\Psi\otimes\Psi\rangle\nonumber\\
		&&=\sum_{12}C_{\alpha_1\beta_1\gamma_1\lambda_1}C_{\alpha_2\beta_2\gamma_2\lambda_2}C^*_{\alpha_2\beta_1\gamma_2\lambda_1}C^*_{\alpha_1\beta_2\gamma_1\lambda_2}.~~
\end{eqnarray}
On the other hand, the density matrix of the whole system (the original system) is $\rho\equiv|\Psi\rangle\langle\Psi|=\sum_{12}C_{\alpha_1\beta_1\gamma_1\lambda_1}C^*_{\alpha_2\beta_2\gamma_2\lambda_2}|\alpha_1\rangle|\beta_1\rangle|\gamma_1\rangle|\lambda_1\rangle\langle\lambda_2|\langle\gamma_2|\langle\beta_2|\langle\alpha_2|$. Therefore, it is readily to get 
\begin{eqnarray}\label{rho2}
	{\rm Tr}(\rho_A^2)
	=\sum_{12}C_{\alpha_1\beta_1\gamma_1\lambda_1}C_{\alpha_2\beta_2\gamma_2\lambda_2}C^*_{\alpha_2\beta_1\gamma_2\lambda_1}C^*_{\alpha_1\beta_2\gamma_1\lambda_2}. 
\end{eqnarray}
We observe that Eq.\eqref{s2} and Eq.\eqref{rho2} are definitely identical. 
Therefore, we can transform the computation of $S_2$ to the computation of the expectation value of the swapping operator as, 
\be\label{link2}
S_2=-\ln {\rm Tr}(\rho_A^2)=-\ln\langle S^{(2)}_{\rm wap}\rangle.
\ee

\subsection{Second-order Renyi entropy $S_2$ for $n$ multiple disjoint intervals}
The above procedure for computing the Renyi entropy of two disjoint intervals can be generically extended to the subsystems with $n$ multiple disjoint intervals. Assume that subsystem $A$ contains $n$ disjoint regions $a_1, a_2, \cdots, a_n$, while the complement part $B$ contains also $n$ disjoint regions $b_1, b_2, \cdots, b_n$. Note that the regions $\{a_i\}$ and $\{b_i\}$ should be alternatingly arranged like $\{a_1b_1a_2b_2\cdots a_nb_n\}$ in order to form the disjoint intervals for the subsystem $A$, i.e, $A=a_1\cup a_2\cup\cdots \cup a_n$ while the the complement part $B=b_1\cup b_2\cup\cdots \cup b_n$. The composite state now is $|\Psi\rangle\otimes|\Psi\rangle$ with $|\Psi\rangle=\sum_{\bf ab}C_{a_1b_1a_2b_2\cdots a_nb_n}|a_1\rangle|b_1\rangle|a_2\rangle|b_2\rangle\cdots|a_n\rangle|b_n\rangle$, where $\sum_{\bf ab}\equiv\sum_{\substack{a_1,a_2,\cdots,a_n\\b_1,b_2,\cdots,b_n}}$. The swapping operator $S^{(2)}_{\rm wap}$ now similarly exchanges the states only in the subsystem $A$ in both $|\Psi\rangle$'s as in Eq.\eqref{swap1}. The rule is as follows,
\begin{widetext}
\begin{eqnarray}
	&&S^{(2)}_{\rm wap}|\Psi\rangle\otimes|\Psi\rangle\nonumber\\
	&&=S^{(2)}_{\rm wap}\left(\sum_{\textbf{ab}}C_{a_1b_1a_2b_2\cdots a_nb_n}|a_1\rangle|b_1\rangle|a_2\rangle|b_2\rangle\cdots|a_n\rangle|b_n\rangle\right)\otimes\left(\sum_{\textbf{cd}}C_{c_1d_1c_2d_2\cdots c_nd_n}|c_1\rangle|d_1\rangle|c_2\rangle|d_2\rangle\cdots|c_n\rangle|d_n\rangle\right)\nonumber\\
	&&=\sum_{\textbf{abcd}}C_{a_1b_1a_2b_2\cdots a_nb_n}C_{c_1d_1c_2d_2\cdots c_nd_n}|c_1\rangle|b_1\rangle|c_2\rangle|b_2\rangle\cdots|c_n\rangle|b_n\rangle\otimes|a_1\rangle|d_1\rangle|a_2\rangle|d_2\rangle\cdots|a_n\rangle|d_n\rangle,
\end{eqnarray}
\end{widetext}
in which $\{|a_i\rangle\}$ are the states in the subsystem $A$ from the first $|\Psi\rangle$ while $\{|c_i\rangle\}$ are the states in the subsystem $A$ from the copied $|\Psi\rangle$. The swapping operator exchanges $\{|a_i\rangle\}\leftrightarrow\{|c_i\rangle\}$. After some computations, it is readily to get that the 2nd-order Renyi entropy $S_2$ of the multiple $n$ disjoint intervals can still be obtained from the expectation value of the swapping operator, i.e, the Eq.\eqref{link2} holds as well for $n$ multiple  disjoint intervals. The detailed proof can be found in the \hyperref[appB]{Appendix B}.

\subsection{Generic $m$-th order Renyi entropy $S_m$ for $n$ multiple disjoint intervals}

\begin{figure}[h]
	\centering
    \includegraphics[trim=4cm 1.2cm 7.2cm 0.7cm, clip=true, scale=0.8]{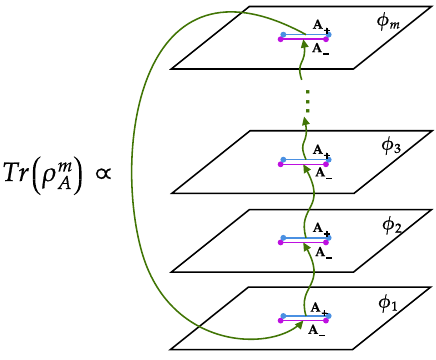}
	\caption{Sketchy picture to illustrate the path integral of ${\rm Tr}(\rho_A^m)$ from the replica trick in quantum field theory. }
	\label{replica}
\end{figure}
For completeness, we further extend the computation of the Renyi entropy to the general $m$-th order, for $n$ multiple disjoint intervals. It is illuminating to see that the swapping operation have some similarities to the replica trick in the quantum field theory \cite{nishioka2009holographic}. In quantum field theory, people always compute the ${\rm Tr}(\rho_A^m)$ by copying the path integral into $m$ sheets, sketched in Fig.\ref{replica}. The interval $A$ is virtually cut into two parts $A_+$ and $A_-$. When doing the path integral, one identifies the field $\phi(A_+)$ on the $k$-th sheet to the $\phi(A_-)$ on the $(k+1)$-st sheet, i.e., $\phi_k(A_+)=\phi_{k+1}(A_-)$, as the short green arrows indicate. The final trace is denoted by the outer big green arrows from $\phi_m(A_+)$ to $\phi_1(A_-)$. 

Inspired from the above replica trick, we speculate that if we want to extend the computation of the Renyi entropy to the general $m$-th order from the swapping operation, we may also need to copy the states $m$ times. 
In this case we need to have $m$ copies of the state $|\Psi\rangle$, i.e., $\otimes_{j=1}^m|\Psi^j\rangle$ where $|\Psi^j\rangle$ is exactly identical to $|\Psi\rangle$. The index $j$ is only used to distinguish the number of $|\Psi\rangle$. Illuminated by the boundary condition in the replica trick, i.e., $\phi_k(A_+)=\phi_{k+1}(A_-)$, here in the swapping operation we need to define a new swapping operator $S^{(m)}_{\rm wap}$, which acts on $\otimes_{j=1}^m|\Psi^j\rangle$ by replacing the states in subsystem $A$ in the $j$-th copied state $|\Psi^j\rangle$ with the states in subsystem $A$ in the $(j+1)$-th copy $|\Psi^{j+1}\rangle$. Note that the states in the subsystem $A$ in the final $m$-th copy $|\Psi^m\rangle$ will be replaced by the subsystem $A$'s states in the first copy $|\Psi^1\rangle$, which resembles the outer big arrow in the Fig.\ref{replica}. The rule is like,
\begin{eqnarray}
	&&{S}^{(m)}_{\rm wap}\otimes_{j=1}^m|\Psi^j\rangle\nno\\
	&&~~~={S}^{(m)}_{\rm wap}\otimes_{j=1}^m[\sum_{\textbf{a}^j\textbf{b}^j}C_{a_1^jb_1^j\cdots a_n^jb_n^j}\otimes_{i=1}^n(|a_i^j\rangle|b_i^j\rangle)]\nno\\
	&&~~~=\otimes_{j=1}^{m-1}[\sum_{\textbf{a}^j\textbf{b}^j}C_{a_1^jb_1^j\cdots a_n^jb_n^j}\otimes_{i=1}^n(|a_i^{j+1}\rangle|b_i^j\rangle)]\nno\\
	&&~~~~~~\otimes[\sum_{\textbf{a}^m\textbf{b}^m}C_{a_1^mb_1^m\cdots a_n^mb_n^m}\otimes_{k=1}^n(|a_{k}^1\rangle|b_{k}^m\rangle)].
\end{eqnarray}

Obviously as $m=2$, the swapping operator $S^{(m)}_{\rm wap}$ returns to $S^{(2)}_{\rm wap}$ from the definition.  After some tedious computations, we found that the $m$-th order Renyi entropy for $n$ multiple disjoint intervals can be computed similarly as that in Eq.\eqref{link2}, i.e, 
\begin{equation}\label{link3}
	S_{m}=\frac{1}{1-m}\ln(\langle S_{\rm wap}^{(m)}\rangle).
\end{equation}
The detailed proof can be found in the \hyperref[appC]{Appendix C}. The above procedure implies that there is indeed resemblance between the replica trick in quantum field theory and the swapping operation as we defined. 




\begin{figure*}[t]
	\centering
	\includegraphics[trim=14.cm 27.cm 15.cm 29.cm, clip=true, scale=0.2]{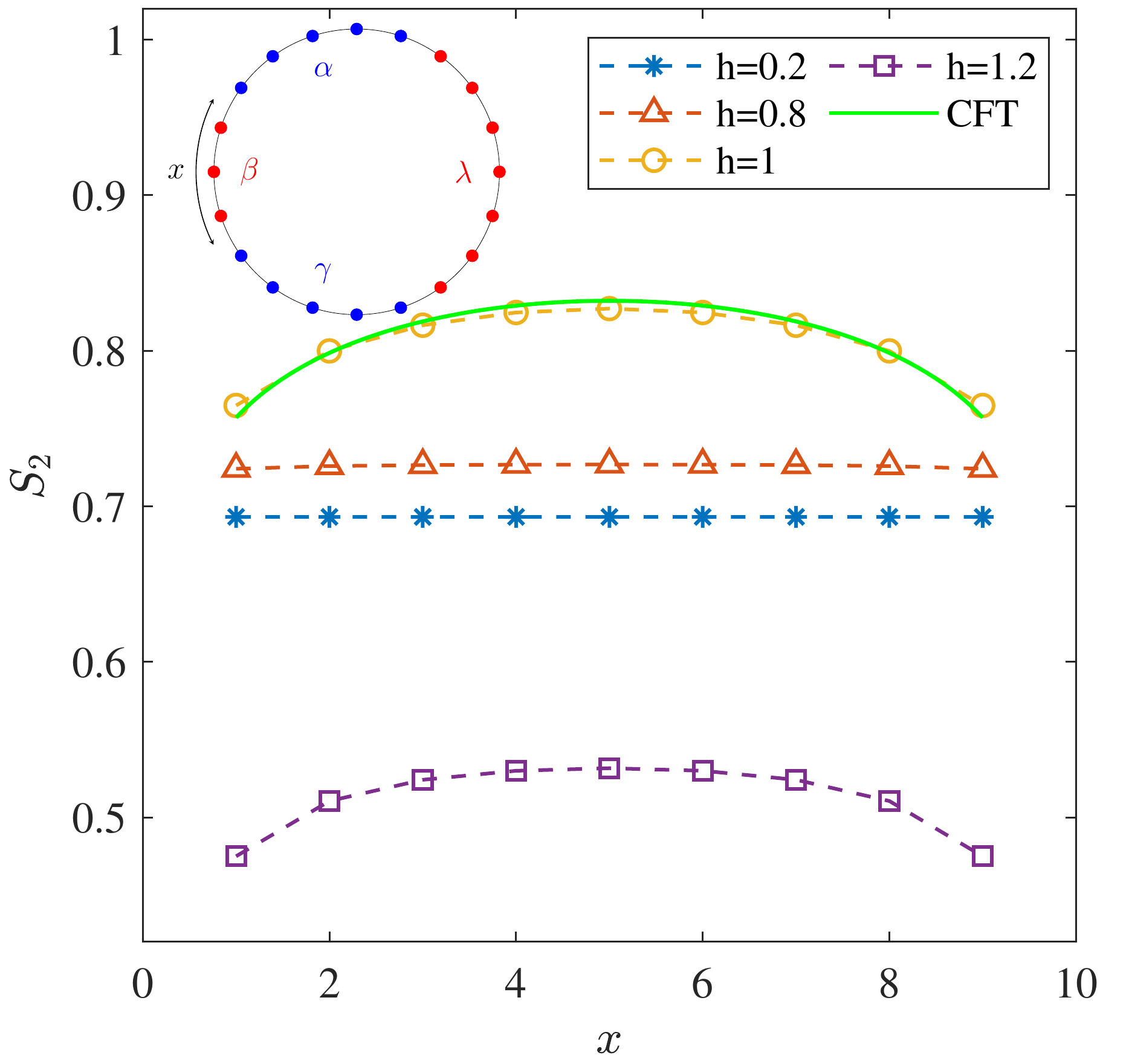}
	\includegraphics[trim=14.cm 27.cm 15.cm 29.cm, clip=true, scale=0.2]{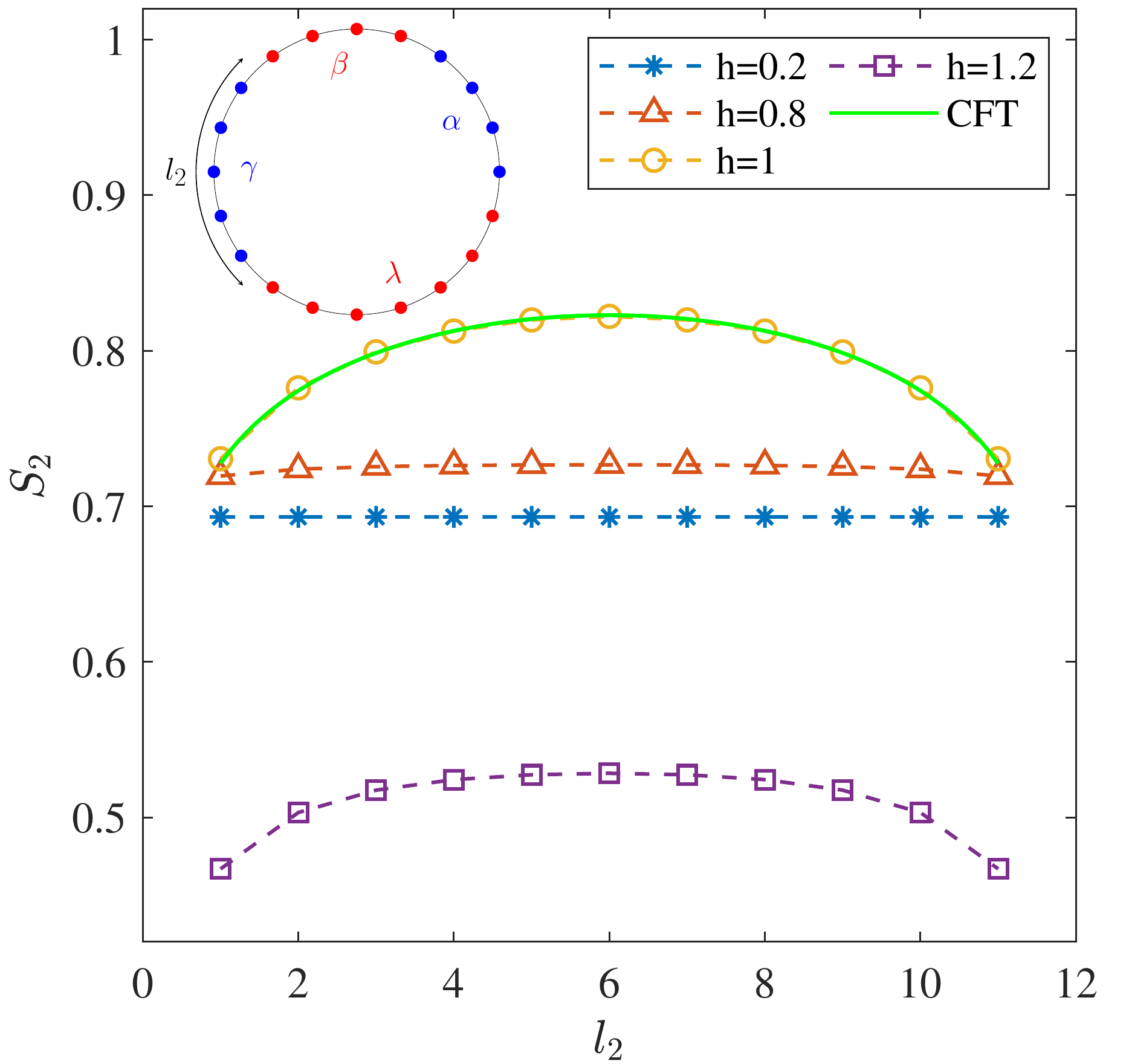}
	\caption{(Left) Renyi entropy $S_2$ for two disjoint intervals against the separation size $x$, with various strengths of the magnetic field $h$; (Right) Renyi entropy $S_2$ for two disjoint intervals against the size of $\gamma$ region $l_2$, with various strengths of the magnetic field $h$. For both panels, the upper-left circles are the sketchy pictures to illustrate the arrangements of the subsystems. The green lines are from the analytical fitting from conformal field theory while the dash-dotted lines are from the results of the swapping operation. }
	\label{interval}
\end{figure*}

\section{Applications in the Ising Model}
We will use the simple and paradigmatic Ising model to demonstrate the above strategy for computing the Renyi entropy. The model is the well-known one dimensional TFQIM with the Hamiltonian,
\begin{equation}\label{TFQIM}
	H=-J\sum_{i=1}^N(\sigma_i^z\sigma_{i+1}^z+h\sigma_i^x),
\end{equation}
in which $N$ is the number of sites, $\sigma_i^z$ and $\sigma_i^x$ are the $z$ and $x$-components of the Pauli matrix at the $i$-th site, $J$ represents the coupling strength between the nearest-neighbor sites while $h$ denotes the strength of the transverse magnetic field. We will work with the periodic boundary conditions for convenience,\footnote{It should be stressed that the procedure for computing the Renyi entropy with swapping operator in the preceding section is not limited to the periodic boundary conditions. From the theory, we can see that it will also work for other boundary conditions.} therefore, the $(N+1)$-th site is equivalent to the first site, i.e., $\vec{\sigma}_{N+1}=\vec{\sigma}_1$. 


One should note that the above method for computing Renyi entropy is viable for various states, not limited to the ground states. But for simplicity of the demonstration, we will work in the ground states to compute the second order Renyi entropy $S_2$, with two, three and four disjoint intervals. These demonstrations are enough to verify the formula Eq.\eqref{link3}. For higher order Renyi entropy the demonstration will be straightforward, and we leave it for future work. 


For a finite spin system, the dimension of its Hamiltonian matrix will grow exponentially with the size of the system. Considering the eigenvalue equation,
\begin{equation}
	H|\Psi_i\rangle=E_i|\Psi_i\rangle
\end{equation}
where $E_i$ and $|\Psi_i\rangle$ are respectively the $i$-th energy eigenvalue and eigenvector. In the TFQIM model \eqref{TFQIM}, we have taken up to $N=24$ sites. The Hamiltonian matrices in this model are typically sparse, therefore, we can diagonalize the Hamiltonian matrix directly with the dimension up to $10^8\times 10^8$. This capacity can be readily realized on a personal computer. We use the singular value decomposition (SVD) to diagonalize the Hamiltonian directly, and then find the eigenstates of $H$. Consequently, the expectation value of the swapping operator $\langle S_{\rm wap}^{(2)}\rangle$ can be computed exactly. 



\begin{figure*}[t]
	\centering
	\includegraphics[trim=14.cm 27.cm 15.cm 29.cm, clip=true, scale=0.2]{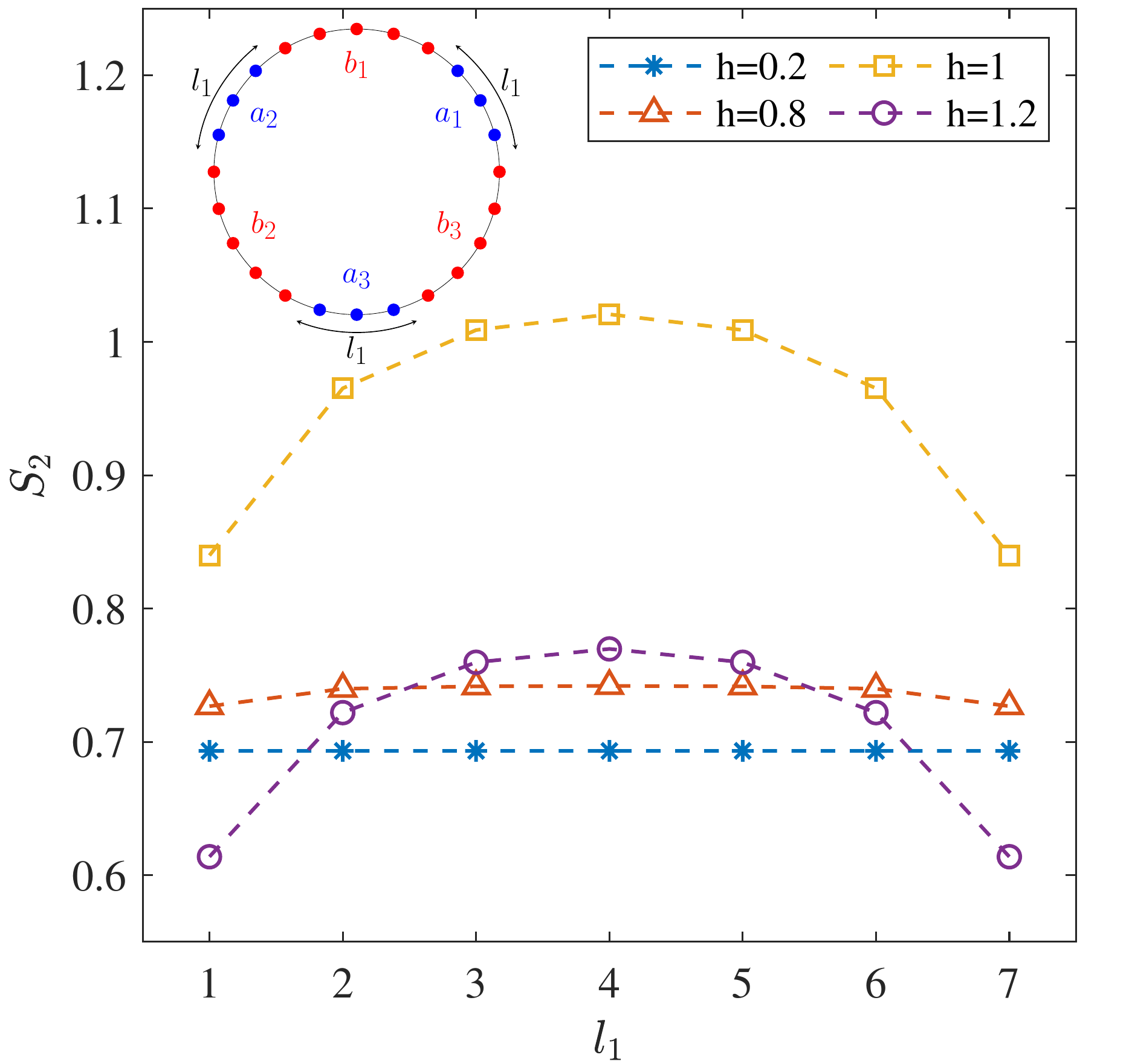}
	\includegraphics[trim=14.cm 27.cm 15.cm 29.cm, clip=true, scale=0.2]{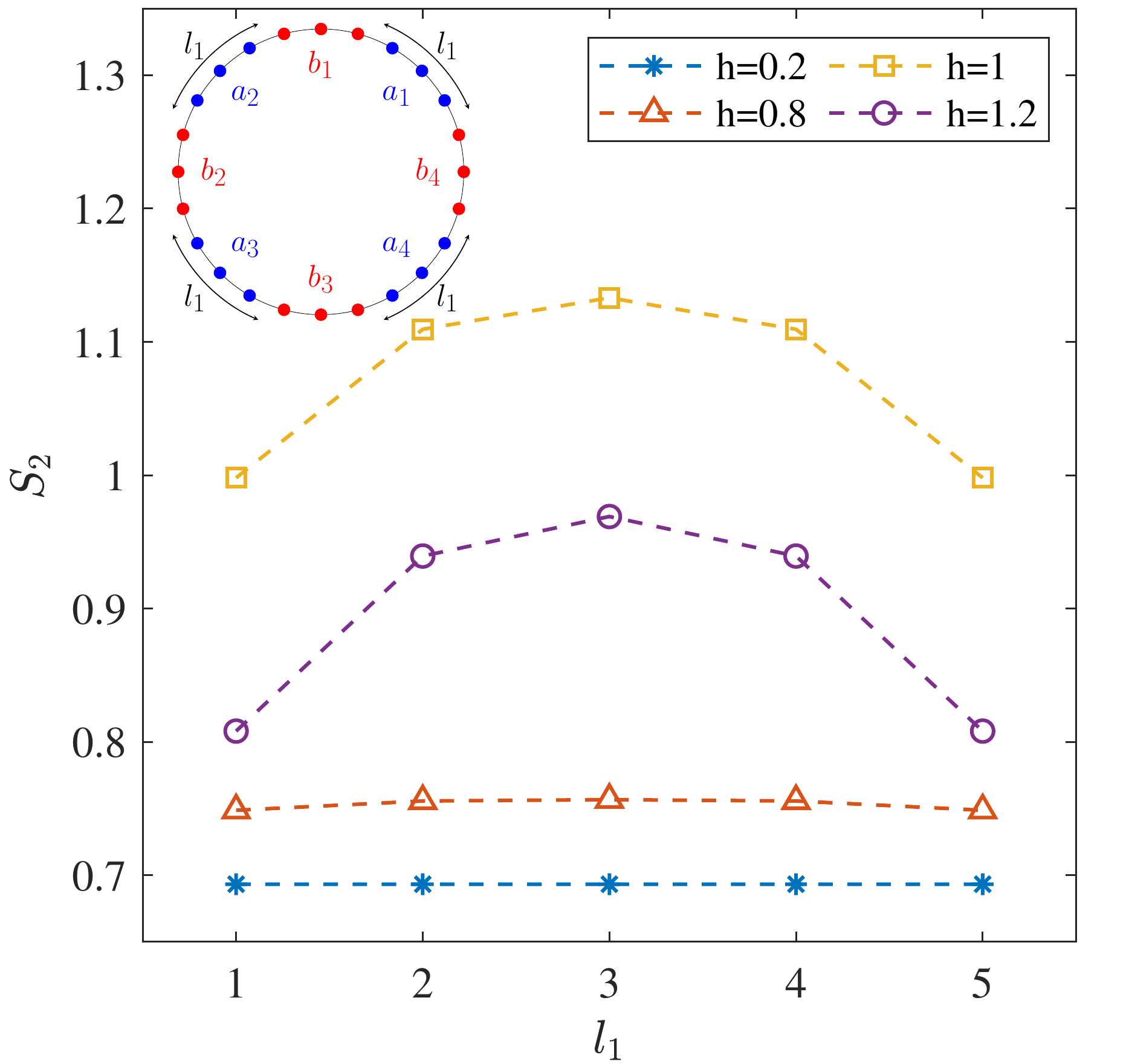}
	\caption{(Left) Renyi entropy $S_2$ for three disjoint intervals $A=a_1\cup a_2\cup a_3$ against the size of $a_{1,2,3}$, i.e., $l_1$; (Right) Renyi entropy $S_2$ for four disjoint intervals $A=a_1\cup a_2\cup a_3\cup a_4$ against the size of $a_{1,2,3,4}$, i.e., $l_1$;  }
	\label{b6}
\end{figure*}

\subsection{Renyi entropy $S_2$ with two disjoint intervals}
We first consider the simplest case of the second order Renyi entropy $S_2$ with two disjoint intervals.  The whole system with the periodic boundary conditions is divided into four subregions, see left panel of Fig.\ref{interval} for illustration. The circle on the upper-left corner has $20$ cites and is divided into $\alpha, \beta, \gamma$ and $\lambda$ subregions. Subsystem $A$ contains two disjoint intervals, i.e, $A=\alpha\cup\gamma$ while the complement is $B=\beta\cup\lambda$. First, we fix the size of each region $\alpha$ and $\gamma$ to be five sites and vary the separation $x$ between these two regions. Because of the periodic boundary conditions, the size of $\beta$ or $\lambda$ play the role as the separation $x$. Specifically, we set the size of the region $\beta$ to be the separation $x$. Therefore, there will a parity symmetry of the Renyi entropy under $x\leftrightarrow (10-x)$. The relation between Renyi entropy $S_2$ and the separation $x$ is exhibited in the left panel of Fig.\ref{interval}, with various magnetic field strength $h$. Obviously, there is a symmetry of $S_2$ under the parity $x\leftrightarrow (10-x)$ due to the periodic boundary conditions.  As $h=0.2$ which is close to zero, the entropy tends to constant $\ln 2$ and depends less on the separation $x$, since in this case the system contains two degenerated ground states $\lvert\uparrow\cdots\uparrow\rangle$ and $\lvert\downarrow\cdots\downarrow\rangle$.  As $h$ increases, the entropy increases as well until it reaches its maximum at the critical magnetic field $h_c=1$. In this case, entropy increases with $x$ until $x=5$. The system at the critical point is in conform field theory. The second Renyi entropy $S_2$ can be analytically computed from the four-point correlation functions \cite{calabrese2011entanglement,furukawa102mutual},
 \begin{equation}
	\text{Tr}\rho_A^m\!\propto\!\left(\frac{x_{31}x_{42}}{x_{21}x_{32}x_{43}x_{41}}\right)^{2\Delta_m}\!\!\left(\frac{\bar{x}_{31}\bar{x}_{42}}{\bar{x}_{21}\bar{x}_{32}\bar{x}_{43}\bar{x}_{41}}\right)^{2\bar{\Delta}_m}\!\!F_m(x,\bar{x})
	\end{equation}
 where $x_{ab}\equiv\frac{L}{\pi}\sin\frac{\pi(x_b-x_a)}{L}$ with $L$ the length of the chain; \footnote{We have defined $\alpha=[x_1, x_2], \beta=[x_2, x_3], \gamma=[x_3, x_4]$ and $\lambda=[x_4, x_1]$. } the conformal dimension $\Delta_m=\bar{\Delta}_m=\frac{c}{24}(m-\frac{1}{m})$ with the central charge $c=\frac{1}{2}$; the cross ratio $x=\frac{x_{21}x_{43}}{x_{31}x_{42}}$ and the function $F_m$ with $m=2$ is
\begin{eqnarray}
	&&F_2(x)=\frac{1}{\sqrt{2}}\left\{\left[\frac{(1+\sqrt{x})(1+\sqrt{1-x})}{2}\right]^{1/2}\right.\nonumber\\
	&&~~~~\left.+x^{1/4}+[(1-x)x]^{1/4}+(1-x)^{1/4}\right\}^{1/2}
\end{eqnarray}
The green line in the left panel of Fig.\ref{interval} is the analytical fitting from the conformal field theory as $h=1$. We can see that the results from conformal field theory is consistent with our results from the swapping operation. 

As the magnetic field exceeds the critical value, system will enter the paramagnetic phase in which different sites will be less correlated. Therefore, for instance of $h=1.2$ we see that the entropy is lower than those at the critical case. Its dependence on the separation $x$ is similar to that of critical point, but since it is away from the conformal critical case, we are lack of the analytical formula to depict it. It is expected that as $h\to\infty$, the entropy will vanish finally since spins on different sites all point to $x$-direction, which are in separable states.   

	

Secondly, we fix the size of regions $\alpha$ and $\beta$ to be four, and then to vary the size of $\gamma$ and $\lambda$. Note that the size of the whole system is still $20$. We denote the size of $\gamma$ as $l_2$. The relationship between Renyi entropy $S_2$ and the size of $\gamma$ is shown in the right panel of Fig.\ref{interval}. Due to the boundary condition, there is a parity symmetry $l_2\leftrightarrow (12-l_2)$ of the entropy. As the magnetic strength is as small as $h=0.2$, the entropy $S_2\approx\ln2$ and depends less on the size of $\gamma$. The reason is similar to that in the preceding paragraph. As $h$ approaches the critical point $h_c=1$, the system possesses conformal symmetry which can be analytically computed. We can see that the green line (from conformal field theory) is consistent with our results from the swapping operations. As $h$ goes beyond $h=1$, the system will enter paramagnetic phase and the entropy will decrease as we have discussed before. 




\subsection{Renyi entropy $S_2$ with three and four disjoint intervals}
Our methods can be readily extended to Renyi entropy with more subregions, such as with three and four subregions, see Fig.\ref{b6}. In this case, we choose the whole system with 24 sites. The subsystem $A=\cup_i a_i$ with $i=1, 2, 3$ for three disjoint intervals (in left panel of Fig.\ref{b6}) or $i=1, 2, 3, 4$ for four disjoint intervals (in right panel of Fig.\ref{b6}). The arrangements of these subregions are shown in the upper-left circles in both panels of Fig.\ref{b6}. We equally partition the subregions $\{a_i\}$ and $\{b_i\}$, and then we simultaneously change the size of the $a_i$ size $l_1$. In the left panel of Fig.\ref{b6}, $l_1\in[1,7]$. From the periodic boundary conditions, there is obviously a parity symmetry $l_1\leftrightarrow (8-l_1)$ which is clear from the plot. In the right panel of Fig.\ref{b6}, $l_2\in[1,5]$. And there is a parity symmetry $l_1\leftrightarrow (6-l_1)$ which is clear from the plot. From both panels, we see that as $h$ is as small as $h=0.2$ the Renyi entropy is roughly constant $\ln2$. And as $h=1$ the Renyi entropy becomes maximum, then decrease as $h$ goes beyond the critical value $h=1$. Since there is no closed formula for three and four disjoint intervals in conformal field theory, we did not fit the Renyi entropy at the critical point as $h=1$. However, our methods with the swapping operation can apply in Renyi entropy with three and four disjoint intervals, which is a big advantage.


\section{Conclusions and Discussions}
We proposed a general approach to compute the Renyi entropy $S_m$ with generally multiple disjoint intervals from the swapping operations. This proposal was based on our observation of the resemblance between the replica trick in quantum field theory with the swapping operations. We found that the computation of $S_m$ can be transformed to evaluate the expectation values of the swapping operator. In order to demonstrate our theory, we studied the second Renyi entropy $S_2$ in the one-dimensional TFQIM with two, three and four disjoint intervals in the ground states. We found that by adjusting the transverse magnetic field, our method can compute the Renyi entropy in or out-of the critical regime of the phase transition. In particular, at the critical point of the system, our results are consistent with those from the conformal field theory.  Therefore, our approach to Renyi entropy may shed light on computing Renyi entropy with multiple intervals. 
It should be stressed that our methods for computing the Renyi entropy is not limited to the ground states, it can also be applied in dynamical cases. Moreover, it can also be applied in higher dimensional systems. We leave it for future work. 



\section*{Acknowledgements}
This work was partially supported by the National Natural Science Foundation of China (Grants No.12175008).

\normalem
\bibliographystyle{unsrt}
\bibliography{ref1.bib}

\begin{widetext}
\appendix
\setcounter{equation}{0}
\setcounter{figure}{0}
\setcounter{table}{0}
\setcounter{section}{0}
\renewcommand{\theequation}{S.\arabic{equation}}
\renewcommand{\thefigure}{S.\arabic{figure}}
\newpage

\begin{center}
--- {\bf Supplemental Materials} ---
\end{center}

\section{Detailed proof of the 2nd-order Renyi entropy with two disjoint intervals}
\label{appA}
In this part, we will show the computational details of the second order Renyi entropy $S_2=-\ln {\rm Tr}(\rho_A^2)$ when the system is divided into an arbitrary number of even parts and a subsystem A contains half the number of regions. Due to the fact that the various within the subsystems are not interconnected, in a one dimensional system, the areas belonging to different subsystem will appear alternately. First, let's consider dividing the system into four regions. Regions $\alpha$ and $\gamma$ belong to subsystem A, while regions $\beta$ and $\lambda$ belong to system B. We use $|\Psi\rangle$ to represent the state of this system.

\begin{equation}
	|\Psi\rangle=\sum_{\alpha\beta\gamma\lambda}C_{\alpha\beta\gamma\lambda}|\alpha\rangle|\beta\rangle|\gamma\rangle|\lambda\rangle
\end{equation}

For calculating $Tr(\rho_A^2)$, one can structure the second order swapping operator $S_{\rm wap}^{(2)}$ acting on the state of two copies of system. It will cut the two states of system, recombine and paste them. 

\begin{eqnarray}
	S^{(2)}_{\rm wap}|\Psi\rangle\otimes|\Psi\rangle&=&S^{(2)}_{\rm wap}\left(\sum_1 C_{\alpha_1\beta_1\gamma_1\lambda_1}|{\alpha_1}\rangle|\beta_1\rangle|{\gamma_1}\rangle|\lambda_1\rangle\right)\otimes\left(\sum_2 C_{\alpha_2\beta_2\gamma_2\lambda_2}|{\alpha_2}\rangle|\beta_2\rangle|{\gamma_2}\rangle|\lambda_2\rangle\right)\nonumber\\
	&=&\sum_{12}C_{\alpha_1\beta_1\gamma_1\lambda_1}C_{\alpha_2\beta_2\gamma_2\lambda_2}|{\alpha_2}\rangle|\beta_1\rangle|{\gamma_2}\rangle|\lambda_1\rangle\otimes|{\alpha_1}\rangle|\beta_2\rangle|{\gamma_1}\rangle|\lambda_2\rangle
\end{eqnarray}
where we wrote $\sum_1$ as $\sum_{\alpha_1\beta_1\gamma_1\lambda_1}$ for abbreviation.
The expectation of swapping operator can be calculated as

\begin{eqnarray}
	\langle S^{(2)}_{\rm wap}\rangle&\equiv&\langle\Psi\otimes\Psi|S^{(2)}_{\rm wap}|\Psi\otimes\Psi\rangle\nonumber\\
	&=&\sum_{1234}C^*_{\alpha_4\beta_4\gamma_4\lambda_4}C^*_{\alpha_3\beta_3\gamma_3\lambda_3}C_{\alpha_1\beta_1\gamma_1\lambda_1}C_{\alpha_2\beta_2\gamma_2\lambda_2}\times\nonumber\\
	&&\langle\lambda_4|\langle\gamma_4|\langle\beta_4|\langle\alpha_4|\langle\lambda_3|\langle\gamma_3|\langle\beta_3|\langle\alpha_3|S^{(2)}_{\rm wap}|\alpha_1\rangle|\beta_1\rangle|\gamma_1\rangle|\lambda_1\rangle|\alpha_2\rangle|\beta_2\rangle|\gamma_2\rangle|\lambda_2\rangle\nonumber\\
	&=&\sum_{1234}C_{\alpha_1\beta_1\gamma_1\lambda_1}C_{\alpha_2\beta_2\gamma_2\lambda_2}C^*_{\alpha_3\beta_3\gamma_3\lambda_3}C^*_{\alpha_4\beta_4\gamma_4\lambda_4}\times\nonumber\\
	&&\langle\lambda_4|\langle\gamma_4|\langle\beta_4|\langle\alpha_4|\langle\lambda_3|\langle\gamma_3|\langle\beta_3|\langle\alpha_3|\alpha_2\rangle|\beta_1\rangle|\gamma_2\rangle|\lambda_1\rangle|\alpha_1\rangle|\beta_2\rangle|\gamma_1\rangle|\lambda_2\rangle\nonumber\\
	&=&\sum_{1234}C_{\alpha_1\beta_1\gamma_1\lambda_1}C_{\alpha_2\beta_2\gamma_2\lambda_2}C^*_{\alpha_3\beta_3\gamma_3\lambda_3}C^*_{\alpha_4\beta_4\gamma_4\lambda_4}\delta_{\alpha_3\alpha_2}\delta_{\beta_3\beta_1}\delta_{\gamma_3\gamma_2}\delta_{\lambda_3\lambda_1}\delta_{\alpha_4\alpha_1}\delta_{\beta_4\beta_2}\delta_{\gamma_4\gamma_1}\delta_{\lambda_4\lambda_2}\nonumber\\
	&=&\sum_{12}C_{\alpha_1\beta_1\gamma_1\lambda_1}C_{\alpha_2\beta_2\gamma_2\lambda_2}C^*_{\alpha_2\beta_1\gamma_2\lambda_1}C^*_{\alpha_1\beta_2\gamma_1\lambda_2}
\end{eqnarray}

Next, we need to verify that the expectation value of the swapping operator is equal to ${\rm Tr}(\rho_A^2)$. Therefore, we need to provide the density matrix $\rho$.

\begin{equation}
	\rho\equiv|\Psi\rangle\langle\Psi|=\sum_{12}C_{\alpha_1\beta_1\gamma_1\lambda_1}C^*_{\alpha_2\beta_2\gamma_2\lambda_2}|\alpha_1\rangle|\beta_1\rangle|\gamma_1\rangle|\lambda_1\rangle\langle\lambda_2|\langle\gamma_2|\langle\beta_2|\langle\alpha_2|
\end{equation}

And the reduced density matrix $\rho_A$ is obtained by taking the trace of $\rho$ over the subsystem B.

\begin{eqnarray}
	\rho_A&\equiv& {\rm Tr}_B(\rho)=\sum_B\langle B|\rho|B\rangle=\sum_{B_3}\langle\lambda_3|\langle\beta_3|\rho|\beta_3\rangle|\lambda_3\rangle\nonumber\\
	&=&\sum_{12B_3}C_{\alpha_1\beta_1\gamma_1\lambda_1}C^*_{\alpha_2\beta_2\gamma_2\lambda_2}\langle\lambda_3|\langle\beta_3|\alpha_1\rangle|\beta_1\rangle|\gamma_1\rangle|\lambda_1\rangle\langle\lambda_2|\langle\gamma_2|\langle\beta_2|\langle\alpha_2|\beta_3\rangle|\lambda_3\rangle\nonumber\\
	&=&\sum_{12B_3}C_{\alpha_1\beta_1\gamma_1\lambda_1}C^*_{\alpha_2\beta_2\gamma_2\lambda_2}\delta_{\lambda_3\lambda_1}\delta_{\beta_3\beta_1}\delta_{\beta_3\beta_2}\delta_{\lambda_3\lambda_2}|\alpha_1\rangle|\gamma_1\rangle\langle\gamma_2|\langle\alpha_2|\nonumber\\
	&=&\sum_{1A_2B_3}C_{\alpha_1\beta_1\gamma_1\lambda_1}C^*_{\alpha_2\beta_3\gamma_2\lambda_3}\delta_{\lambda_3\lambda_1}\delta_{\beta_3\beta_1}|\alpha_1\rangle|\gamma_1\rangle\langle\gamma_2|\langle\alpha_2|\nonumber\\
	&=&\sum_{1A_2}C_{\alpha_1\beta_1\gamma_1\lambda_1}C^*_{\alpha_2\beta_1\gamma_2\lambda_1}|\alpha_1\rangle|\gamma_1\rangle\langle\gamma_2|\langle\alpha_2|
\end{eqnarray}

Then calculate the square of $\rho_A$.

\begin{eqnarray}
	\rho_A^2&=&\sum_{1A_3}C_{\alpha_1\beta_1\gamma_1\lambda_1}C^*_{\alpha_3\beta_1\gamma_3\lambda_1}|\alpha_1\rangle|\gamma_1\rangle\langle\gamma_3|\langle\alpha_3|\sum_{2A_4}C_{\alpha_2\beta_2\gamma_2\lambda_2}C^*_{\alpha_4\beta_2\gamma_4\lambda_2}|\alpha_2\rangle|\gamma_2\rangle\langle\gamma_4|\langle\alpha_4|\nonumber\\
	&=&\sum_{12A_3A_4}C_{\alpha_1\beta_1\gamma_1\lambda_1}C_{\alpha_2\beta_2\gamma_2\lambda_2}C^*_{\alpha_3\beta_1\gamma_3\lambda_1}C^*_{\alpha_4\beta_2\gamma_4\lambda_2}\delta_{\gamma_3\gamma_2}\delta_{\alpha_3\alpha_2}|\alpha_1\rangle|\gamma_1\rangle\langle\gamma_4|\langle\alpha_4|\nonumber\\
	&=&\sum_{12A_4}C_{\alpha_1\beta_1\gamma_1\lambda_1}C_{\alpha_2\beta_2\gamma_2\lambda_2}C^*_{\alpha_2\beta_1\gamma_2\lambda_1}C^*_{\alpha_4\beta_2\gamma_4\lambda_2}|\alpha_1\rangle|\gamma_1\rangle\langle\gamma_4|\langle\alpha_4|
\end{eqnarray}

Finally, taking the trace of $\rho_A^2$, we can find that the result is consistent with the expectation value of swapping operator.

\begin{eqnarray}
	{\rm Tr}(\rho_A^2)&=&\sum_{12A_4}C_{\alpha_1\beta_1\gamma_1\lambda_1}C_{\alpha_2\beta_2\gamma_2\lambda_2}C^*_{\alpha_2\beta_1\gamma_2\lambda_1}C^*_{\alpha_4\beta_2\gamma_4\lambda_2}\delta_{\alpha_1\alpha_4}\delta_{\gamma_1\gamma_4}\nonumber\\
	&=&\sum_{12}C_{\alpha_1\beta_1\gamma_1\lambda_1}C_{\alpha_2\beta_2\gamma_2\lambda_2}C^*_{\alpha_2\beta_1\gamma_2\lambda_1}C^*_{\alpha_1\beta_2\gamma_1\lambda_2}=\langle S_{\rm wap}^{(2)}\rangle
\end{eqnarray}

Therefore, the connection from swapping operator to two-order Renyi entropy has been built. It allows us to measure the entropy $S_2$ via importance sampling of the swapping operator.

\begin{equation}
	S_2=-\ln\langle S^{(2)}_{\rm wap}\rangle
\end{equation}

\section{Detailed proof of the 2nd-order Renyi entropy with $n$ multiple disjoint intervals}
\label{appB}
Now, we are considering the case where the subsystem contains more disjointed regions. The subsystem includes regions $a_1,a_2,\cdots,a_n$, while subsystem B contains $b_1,b_2,\cdots,b_n$. Since the regins within the same subsystem are not connected, regions $a_i$ and $b_i$ will alternate in the state.

\begin{equation}
	|\Psi\rangle=\sum_{ab}C_{a_1b_1a_2b_2\cdots a_nb_n}|a_1\rangle|b_1\rangle|a_2\rangle|b_2\rangle\cdots|a_n\rangle|b_n\rangle
\end{equation}

The swapping operator $S_{\rm wap}^{(2)}$ defined under this state still involves cutting the two states, exchanging their parts of subsystem A, and pasting them back together.

\begin{eqnarray}
	&&S^{(2)}_{\rm wap}|\Psi\rangle\otimes|\Psi\rangle\nonumber\\
	&=&S^{(2)}_{\rm wap}\left(\sum_{ab}C_{a_1b_1a_2b_2\cdots a_nb_n}|a_1\rangle|b_1\rangle|a_2\rangle|b_2\rangle\cdots|a_n\rangle|b_n\rangle\right)\otimes\left(\sum_{cd}C_{c_1d_1c_2d_2\cdots c_nd_n}|c_1\rangle|d_1\rangle|c_2\rangle|d_2\rangle\cdots|c_n\rangle|d_n\rangle\right)\nonumber\\
	&=&\sum_{abcd}C_{a_1b_1a_2b_2\cdots a_nb_n}C_{c_1d_1c_2d_2\cdots c_nd_n}|c_1\rangle|b_1\rangle|c_2\rangle|b_2\rangle\cdots|c_n\rangle|b_n\rangle\otimes|a_1\rangle|d_1\rangle|a_2\rangle|d_2\rangle\cdots|a_n\rangle|d_n\rangle.
\end{eqnarray}

we also calculate the expectation value of swapping operator, and compare it with ${\rm Tr}(\rho_a^2)$.

\begin{eqnarray}
	\langle S^{(2)}_{\rm wap}\rangle&\equiv&\langle\Psi\otimes\Psi|S^{(2)}_{\rm wap}|\Psi\otimes\Psi\rangle\nonumber\\
	&=&\sum_{\substack{abcd\\a'b'c'd'}}C_{a_1b_1a_2b_2\cdots a_nb_n}C_{c_1d_1c_2d_2\cdots c_nd_n}C_{a'_1b'_1a'_2b'_2\cdots a'_nb'_n}^*C_{c'_1d'_1c'_2d'_2\cdots c'_nd'_n}^*\langle d'_n|\langle c'_n|\cdots\langle d'_2|\langle c'_2|\langle d'_1|\langle c'_1|\nonumber\\
	&&~~~~~~~\langle b'_n|\langle a'_n|\cdots\langle b'_2|\langle a'_2|\langle b'_1|\langle a'_1|S_{\rm wap}^{(2)}|a_1\rangle|b_1\rangle|a_2\rangle|b_2\rangle\cdots|a_n\rangle|b_n\rangle|c_1\rangle|d_1\rangle|c_2\rangle|d_2\rangle\cdots|c_n\rangle|d_n\rangle\nonumber\\
	&=&\sum_{\substack{abcd\\a'b'c'd'}}C_{a_1b_1a_2b_2\cdots a_nb_n}C_{c_1d_1c_2d_2\cdots c_nd_n}C_{a'_1b'_1a'_2b'_2\cdots a'_nb'_n}^*C_{c'_1d'_1c'_2d'_2\cdots c'_nd'_n}^*\langle d'_n|\langle c'_n|\cdots\langle d'_2|\langle c'_2|\langle d'_1|\langle c'_1|\nonumber\\
	&&~~~~~~~\langle b'_n|\langle a'_n|\cdots\langle b'_2|\langle a'_2|\langle b'_1|\langle a'_1|c_1\rangle|b_1\rangle|c_2\rangle|b_2\rangle\cdots|c_n\rangle|b_n\rangle|a_1\rangle|d_1\rangle|a_2\rangle|d_2\rangle\cdots|a_n\rangle|d_n\rangle\nonumber\\
	&=&\sum_{\substack{abcd\\a'b'c'd'}}C_{a_1b_1a_2b_2\cdots a_nb_n}C_{c_1d_1c_2d_2\cdots c_nd_n}C_{a'_1b'_1a'_2b'_2\cdots a'_nb'_n}^*C_{c'_1d'_1c'_2d'_2\cdots c'_nd'_n}^*\times\nonumber\\
	&&~~~~~~~\delta_{c_1a'_1}\delta_{b_1b'_1}\delta_{c_2a'_2}\delta_{b_2b'_2}\cdots\delta_{c_na'_n}\delta_{b_nb'_n}\delta_{a_1c'_1}\delta_{d_1d'_1}\delta_{a_2c'_2}\delta_{d_2d'_2}\cdots\delta_{a_nc'_n}\delta_{d_nd'_n}\nonumber\\
	&=&\sum_{abcd}C_{a_1b_1a_2b_2\cdots a_nb_n}C_{c_1d_1c_2d_2\cdots c_nd_n}C_{c_1b_1c_2b_2\cdots c_nb_n}^*C_{a_1d_1a_2d_2\cdots a_nd_n}^*
\end{eqnarray}

To calculate ${\rm Tr}(\rho_A^2)$, we first need to provide the density matrix $\rho$.

\begin{eqnarray}
	\rho&\equiv&|\Psi\rangle\langle\Psi|\nonumber\\
	&=&\sum_{abcd}C_{a_1b_1a_2b_2\cdots a_nb_n}C^*_{c_1d_1c_2d_2\cdots c_nd_n}|a_1\rangle|b_1\rangle|a_2\rangle|b_2\rangle\cdots|a_n\rangle|b_n\rangle\langle d_n|\langle c_n|\cdots\langle d_2|\langle c_2|\langle d_1|\langle c_1|.~~~~
\end{eqnarray}

And the density matrix $\rho_A$.

\begin{eqnarray}
	\rho_A&\equiv& {\rm Tr}_B(\rho)=\sum_{b'}\langle b'_n|\cdots\langle b'_2|\langle b'_1|\rho|b'_1\rangle|b'_2\rangle\cdots|b'_n\rangle\nonumber\\
	&=&\sum_{abcdb'}C_{a_1b_1a_2b_2\cdots a_nb_n}C_{c_1d_1c_2d_2\cdots c_nd_n}^*\times\nonumber\\
	&&~~~~~~\langle b'_n|\cdots\langle b'_2|\langle b'_1|a_1\rangle|b_1\rangle|a_2\rangle|b_2\rangle\cdots|a_n\rangle|b_n\rangle\langle d_n|\langle c_n|\cdots\langle d_2|\langle c_2|\langle d_1|\langle c_1|b'_1\rangle|b'_2\rangle\cdots|b'_n\rangle\nonumber\\
	&=&\sum_{abcdb'}C_{a_1b_1a_2b_2\cdots a_nb_n}C_{c_1d_1c_2d_2\cdots c_nd_n}^*\delta_{b_1b'_1}\delta_{b_2b'_2}\cdots\delta_{b_nb'_n}\delta_{b'_1d_1}\delta_{b'_2d_2}\cdots\delta_{b'_nd_n}|a_1\rangle|a_2\rangle\cdots|a_n\rangle\langle c_n|\cdots\langle c_2|\langle c_1|\nonumber\\
	&=&\sum_{abc}C_{a_1b_1a_2b_2\cdots a_nb_n}C_{c_1b_1c_2b_2\cdots c_nb_n}^*|a_1\rangle|a_2\rangle\cdots|a_n\rangle\langle c_n|\cdots\langle c_2|\langle c_1|
\end{eqnarray}

Calculating $\rho_A^2$ following

\begin{eqnarray}
	\rho_A^2&=&\sum_{aba'cdc'}C_{a_1b_1a_2b_2\cdots a_nb_n}C_{a'_1b_1a'_2b_2\cdots a'_nb_n}^*C_{c_1d_1c_2d_2\cdots c_nd_n}C_{c'_1d_1c'_2d_2\cdots c'_nd_n}^*\times\nonumber\\
	&&~~~~~~~~|a_1\rangle|a_2\rangle\cdots|a_n\rangle\langle a'_n|\cdots\langle a'_2|\langle a'_1|c_1\rangle|c_2\rangle\cdots|c_n\rangle\langle c'_n|\cdots\langle c'_2|\langle c'_1|\nonumber\\
	&=&\sum_{aba'cdc'}C_{a_1b_1a_2b_2\cdots a_nb_n}C_{a'_1b_1a'_2b_2\cdots a'_nb_n}^*C_{c_1d_1c_2d_2\cdots c_nd_n}C_{c'_1d_1c'_2d_2\cdots c'_nd_n}^*\times\nonumber\\
	&&~~~~~~~~\delta_{a'_1c_1}\delta_{a'_2c_2}\cdots\delta_{a'_nc_n}|a_1\rangle|a_2\rangle\cdots|a_n\rangle\langle c'_n|\cdots\langle c'_2|\langle c'_1|\nonumber\\
	&=&\!\!\sum_{abcdc'}C_{a_1b_1a_2b_2\cdots a_nb_n}C_{c_1b_1c_2b_2\cdots c_nb_n}^*C_{c_1d_1c_2d_2\cdots c_nd_n}C_{c'_1d_1c'_2d_2\cdots c'_nd_n}^*|a_1\rangle|a_2\rangle\cdots|a_n\rangle\langle c'_n|\cdots\langle c'_2|\langle c'_1|. ~~~~~~~
\end{eqnarray}

By calculation, the expectation calue of the swapping operator is consistent with $Tr(\rho_A^2)$. Therefore, the Renyi entropy $S_2$ for a subsystem containing an arbitary number of disconnected regions can still be obtained through the expectation value of the swapping operator.

\begin{eqnarray}
	{\rm Tr}(\rho_A^2)&=&\sum_{abcdc'}C_{a_1b_1a_2b_2\cdots a_nb_n}C_{c_1b_1c_2b_2\cdots c_nb_n}^*C_{c_1d_1c_2d_2\cdots c_nd_n}C_{c'_1d_1c'_2d_2\cdots c'_nd_n}^*\delta_{a_1c'_1}\delta_{a_2c'_2}\cdots\delta_{a_nc'_n}\nonumber\\
	&=&\sum_{abcd}C_{a_1b_1a_2b_2\cdots a_nb_n}C_{c_1d_1c_2d_2\cdots c_nd_n}C_{c_1b_1c_2b_2\cdots c_nb_n}^*C_{a_1d_1a_2d_2\cdots a_nd_n}^*=\langle S_{\rm wap}^{(2)}\rangle
\end{eqnarray}

Then we still get 
\begin{equation}
	S_2=-\ln\langle S^{(2)}_{\rm wap}\rangle
\end{equation}

\section{Detailed proof of the $m$-th order Renyi entropy $S_m$ with $n$ multiple disjoint intervals}
\label{appC}
In this part, we will show the computational details of arbitary order R\'enyi entropy with $n$ intervals. Firstly we set the order number of R\'enyi entropy is $m$

\begin{equation}
	S_{m}=\frac{1}{1-m}\ln\text{Tr}(\rho_A^m)
\end{equation}

We cut the system to two subsystems A and B, while both A and B consist of $n$ parts $a_i$ and $b_i$ ($i=1,2,\cdots,n$). They are interleaved on a one-dimensional system. Specifically, $a_1$ appears at the beginning of the system, followed by $b_1$, then $a_2$, and so on. Under this partition, the wave function of the entire system can be represented as

\begin{equation}
	|\Psi\rangle=\sum_{\textbf{ab}}C_{a_1b_1\cdots a_nb_n}\otimes_{i=1}^n(|a_i\rangle|b_i\rangle)
\end{equation}	
in which $\otimes_{i=1}^n$ is the $n$ times of the tensor product of the states $|a_i\rangle|b_i\rangle$. 
To obtain the value of $\text{Tr}(\rho_A^m)$, one can structure the m-th order swapping operator ${S}_{\rm wap}^{(m)}$ acting the state of $m$ copies of system.

\begin{eqnarray}
	&&{S}^{(m)}_{\rm wap}\otimes_{j=1}^m|\Psi^j\rangle={S}^{(m)}_{\rm wap}\otimes_{j=1}^m[\sum_{\textbf{a}^j\textbf{b}^j}C_{a_1^jb_1^j\cdots a_n^jb_n^j}\otimes_{i=1}^n(|a_i^j\rangle|b_i^j\rangle)]\\
	&&=\otimes_{j=1}^{m-1}[\sum_{\textbf{a}^j\textbf{b}^j}C_{a_1^jb_1^j\cdots a_n^jb_n^j}\otimes_{i=1}^n(|a_i^{j+1}\rangle|b_i^j\rangle)]\otimes[\sum_{\textbf{a}^m\textbf{b}^m}C_{a_1^mb_1^m\cdots a_n^mb_n^m}\otimes_{k=1}^n(|a_{k}^1\rangle|b_{k}^m\rangle)]
\end{eqnarray}

Applying ${S}_{\rm wap}^{(m)}$ to the state of $m$ tensor-product systems, it has the effect of replacing the subsystem $A^{j+1}$ of $(j+1)$-th replica, i.e., $a_i^{j+1}$, with the subsystem $A^j$ of the $j$-th replica. This replacement is done for every two adjacent replicas, and finally, the subsystem $A^1$ of the first replica is replaced with the subsystem $A^m$ of the last replica. Under this definition, the expectation of swapping operator can be calculated as

\begin{eqnarray}
	&&\langle S_{\rm wap}^{(m)}\rangle=\langle\otimes_{j'=1}^m\Psi^{j'}|S_{\rm wap}^{(m)}|\otimes_{j=1}^m\Psi^{j}\rangle\\
	=&&\otimes_{j'=1}^m[\sum_{\textbf{a}'^{j'}\textbf{b}'^{j'}}C^*_{{a'}_1^{j'}{b'}_1^{j'}\cdots {a'}_n^{j'}{b'}_n^{j'}}\otimes_{i'=1}^n(\langle {a'}_{i'}^{j'}|\langle {b'}_{i'}^{j'}|)]\otimes_{j=1}^{m-1}[\sum_{\textbf{a}^j\textbf{b}^j}C_{a_1^jb_1^j\cdots a_n^jb_n^j}\otimes_{i=1}^n(|a_i^{j+1}\rangle|b_i^j\rangle)]\nonumber\\
	&&\otimes[\sum_{\textbf{a}^m\textbf{b}^m}C_{a_1^mb_1^m\cdots a_n^mb_n^m}\otimes_{k=1}^n(|a_k^1\rangle|b_k^m\rangle)]\\
	=&&\prod_{j,j'=1}^{m-1}\sum_{\substack{\textbf{a}^j\textbf{b}^j\\\textbf{a}'^{j'}\textbf{b}'^{j'}}}[C^*_{{a'}_1^{j'}{b'}_1^{j'}\cdots {a'}_n^{j'}{b'}_n^{j'}}C_{a_1^jb_1^j\cdots a_n^jb_n^j}\prod_{i=1}^n(\delta_{{a'}_i^{j'}a_i^{j+1}}\delta_{{b'}_i^{j'}b_i^j})]\nonumber\\
	&&\times\sum_{\substack{\textbf{a}^m\textbf{b}^m\\\textbf{a}'^m\textbf{b}'^m}}[C^*_{{a'}_1^m{b'}_1^m\cdots{a'}_n^m{b'}_n^m}C_{a_1^mb_1^m\cdots a_n^mb_n^m}\prod_{k=1}^n(\delta_{{a'}_k^ma_k^1}\delta_{{b'}_k^mb_k^m})]\\
	=&&\prod_{j=1}^{m-1}\sum_{\textbf{a}^j\textbf{b}^j}[C^*_{a_1^{j+1}b_1^j\cdots a_n^{j+1}b_n^j}C_{a_1^jb_1^j\cdots a_n^jb_n^j}]\times\sum_{\textbf{a}^m\textbf{b}^m}[C^*_{a_1^1b_1^m\cdots a_n^1b_n^m}C_{a_1^mb_1^m\cdots a_n^mb_n^m}]
\end{eqnarray}

To verify $\text{Tr}(\rho_A^m)$ is equal to the expectation value of ${S}_{\rm wap}^{(m)}$, we need to perform calculations from another perspective. First, we present the representation of the density matrix $\rho$

\begin{equation}
	\rho=|\Psi\rangle\langle\Psi|=\sum_{\textbf{abcd}}C_{a_1b_1\cdots a_nb_n}C^*_{c_1d_1\cdots c_nd_n}\otimes_{i=1}^n(|a_i\rangle|b_i\rangle)\otimes_{i'=1}^n(\langle c_{i'}|\langle d_{i'}|)
\end{equation}

The reduced density matrix $\rho_A$ for subsystem A is obtained by tracing out subsystem B from the density matrix $\rho$

\begin{eqnarray}
	&&\rho_A=\text{Tr}_B(\rho)=\sum_{\textbf{b}'}\otimes_{j=1}^n(\langle b'_j|)\rho\otimes_{j'=1}^n(|b'_{j'}\rangle)\\
	=&&\sum_{\textbf{b}'}\sum_{\textbf{abcd}}C_{a_1b_1\cdots a_nb_n}C^*_{c_1d_1\cdots c_nd_n}\otimes_{j=1}^n(\langle b'_j|)\otimes_{i=1}^n(|a_i\rangle|b_i\rangle)\otimes_{i'=1}^n(\langle c_{i'}|\langle d_{i'}|)\otimes_{j'=1}^n(|b'_{j'}\rangle)\\
	=&&\sum_{\textbf{abcdb}'}C_{a_1b_1\cdots a_nb_n}C^*_{c_1d_1\cdots c_nd_n}\prod_{j=1}^n(\delta_{b'_jb_j})\prod_{j'=1}^n(\delta_{d_{j'}b'_{j'}})\otimes_{i=1}^n(|a_i\rangle)\otimes_{i'=1}^n(\langle c_{i'}|)\\
	=&&\sum_{\textbf{abc}}C_{a_1b_1\cdots a_nb_n}C^*_{c_1b_1\cdots c_nb_n}\otimes_{i=1}^n(|a_i\rangle)\otimes_{i'=1}^n(\langle c_{i'}|)
\end{eqnarray}

Next, we calculate the m-th power of reduced density matrix

\begin{eqnarray}
	&&\rho_A^m=\prod_{j=1}^m[\sum_{\textbf{a}^j\textbf{b}^j\textbf{c}^j}C_{a_1^jb_1^j\cdots a_n^jb_n^j}C^*_{c_1^jb_1^j\cdots c_n^jb_n^j}\otimes_{i=1}^n(|a_i^j\rangle)\otimes_{i'=1}^n(\langle c_{i'}^j|)]\\
	=&&\prod_{j=1}^{m-1}[\sum_{\textbf{a}^j\textbf{b}^j\textbf{c}^j}C_{a_1^jb_1^j\cdots a_n^jb_n^j}C^*_{c_1^jb_1^j\cdots c_n^jb_n^j}\prod_{i=1}^n(\delta_{c_i^ja_i^{j+1}})]\nonumber\\
	&&\times\sum_{\textbf{a}^m\textbf{b}^m\textbf{c}^m}[C_{a_1^mb_1^m\cdots a_n^mb_n^m}C^*_{c_1^mb_1^m\cdots c_n^mb_n^m}\otimes_{k=1}^n(|a_k^1\rangle)\otimes_{k'=1}^n(\langle c_{k'}^m|)]\\
	=&&\prod_{j=1}^{m-1}[\sum_{\textbf{a}^j\textbf{b}^j}C_{a_1^jb_1^j\cdots a_n^jb_n^j}C^*_{a_1^{j+1}b_1^j\cdots a_n^{j+1}b_n^j}]\nonumber\\
	&&\times\sum_{\textbf{a}^m\textbf{b}^m\textbf{c}^m}[C_{a_1^mb_1^m\cdots a_n^mb_n^m}C^*_{c_1^mb_1^m\cdots c_n^mb_n^m}\otimes_{k=1}^n(|a_k^1\rangle)\otimes_{k'=1}^n(\langle c_{k'}^m|)]
\end{eqnarray}

Then, take its trace

\begin{eqnarray}
	&&\text{Tr}(\rho_A^m)=\prod_{j=1}^{m-1}[\sum_{\textbf{a}^j\textbf{b}^j}C_{a_1^jb_1^j\cdots a_n^jb_n^j}C^*_{a_1^{j+1}b_1^j\cdots a_n^{j+1}b_n^j}]\times\!\!\!\!\!\sum_{\textbf{a}^m\textbf{b}^m\textbf{c}^m}\!\!\![C_{a_1^mb_1^m\cdots a_n^mb_n^m}C^*_{c_1^mb_1^m\cdots c_n^mb_n^m}\prod_{k=1}^n(\delta_{a_k^1c_k^m})]\\
	&&=\prod_{j=1}^{m-1}[\sum_{\textbf{a}^j\textbf{b}^j}C_{a_1^jb_1^j\cdots a_n^jb_n^j}C^*_{a_1^{j+1}b_1^j\cdots a_n^{j+1}b_n^j}]\times\!\!\!\!\sum_{\textbf{a}^m\textbf{b}^m}\!\![C_{a_1^mb_1^m\cdots a_n^mb_n^m}C^*_{a_1^1b_1^m\cdots a_n^1b_n^m}]=\langle S_{\rm wap}^{(m)}\rangle
\end{eqnarray}

It can be seen that the value of $\text{Tr}(\rho_A^m)$ is consistent with the expectation value of the operator ${S_{\rm wap}^{(m)}}$. Therefore, the relationship between the m-th order R\'enyi entropy and $\langle S_{\rm wap}^{(m)}\rangle$ is

\begin{equation}
	S_{m}=\frac{1}{1-m}\ln(\langle S_{\rm wap}^{(m)}\rangle)
\end{equation}

\end{widetext}
\end{document}